\documentclass[preliminary]{eptcs}
\def\thepage{\arabic{page}}

\usepackage{hyperref}

\usepackage{graphicx}
\usepackage{comment}
\usepackage{wrapfig}
\usepackage[pat,steen,nico]{mydraft}
\usepackage{nico}
\usepackage{tikz}
\usetikzlibrary{arrows,automata,shapes,snakes}
\usepackage{slashbox,hhline,longtable}
\usepackage{thm-restate}
\usepackage{symnash}
\usepackage{enumerate}
\IfFileExists{microtype}{\usepackage{microtype}}{}

\def\refname{References}
\newtheorem{problem}{Problem}

\begin{document}

\def\event{2nd International Workshop on Strategic Reasoning (SR'14). April 5-6, 2014, Grenoble, France}
\title{\hbox to \linewidth{\hss Nash Equilibria in Symmetric Games with Partial Observation\hss}}
\author{
\end{tabular}\begin{tabular}{>{\centering}p{.3\textwidth}>{\centering}p{.3\textwidth}>{\centering}p{.3\textwidth}} 
  Patricia Bouyer & Nicolas Markey & Steen Vester \tabularnewline
\multicolumn{2}{c}{\small LSV, CNRS \& ENS~Cachan, France} & \small DTU, Kgs. Lyngby, 
  Denmark \tabularnewline
\multicolumn{2}{c}{\small \href{mailto:bouyer@lsv.ens-cachan.fr,markey@lsv.ens-cachan.fr}{\texttt{\{bouyer,markey\}@lsv.ens-cachan.fr}}} &
\small \href{mailto:stve@dtu.dk}{stve@dtu.dk}}
\def\titlerunning{Nash Equilibria in Symmetric Game with Partial Observation}
\def\authorrunning{P. Bouyer, N. Markey, S. Vester}

\specialthanks{Part of this work was sponsored by ERC Starting Grant EQualIS
  and by EU FP7 project Cassting.}

\maketitle

\begin{abstract}
  \looseness=-1
  We investigate a model for representing large multiplayer games,
  which satisfy strong symmetry properties. This model is made of
  multiple copies of an arena; each player plays in his own arena,
  and can partially observe what the other players~do. Therefore, this
  game has partial information and symmetry constraints, which
  make the computation of Nash equilibria difficult. We show several
  undecidability results, and for bounded-memory strategies, we
  precisely characterize the complexity of computing pure
  Nash equilibria (for qualitative objectives) in this game model.
\end{abstract}

\section{Introduction}

\paragraph{Multiplayer games.}
\looseness=-1
Games played on graphs have been intensively used in computer science
as a tool to reason about and automatically synthesize interacting
reactive systems~\cite{tark2005-Hen}.
Consider a server granting access to a printer and connected to
several clients. The clients may send requests to the server, and the
server grants access to the printer depending on the requests it
receives. The server could have various strategies: for instance,
never grant access to any client, or always immediately grant access
upon request.  However, it~may also have constraints to satisfy (which
define its winning condition): for instance, that no two clients
should access the printer at the same time, or that any request
must eventually be granted.  A~strategy for the server is then a
policy that it should apply in order to achieve these goals.

Until recently, more focus had been put on the study of purely
antagonistic games (a.k.a.~zero-sum games), which conveniently
represent systems evolving in a (hostile) environment:
the aim of one player is to prevent the other player from achieving
his own objective.

\paragraph{Non-zero-sum games.}
\looseness=-1
Over the last ten years, computer scientists have started considering games
with non-zero-sum objectives: they~allow for conveniently modelling complex
infrastructures where each individual system tries to fulfill its own
objectives, while still being subject to uncontrollable actions of the
surrounding systems. As an example, consider a wireless network in which
several devices try to send data: each device can modulate its transmitting power,
in order to maximize its bandwidth or reduce energy consumption as much as
possible. In~that setting, focusing only on optimal strategies for one single
agent is too narrow. Game-theoreticians have defined and studied many other
solution concepts for such settings, of which Nash equilibrium~\cite{Nas50} is
a prominent~one. A~Nash equilibrium is a strategy profile where no player
can improve the outcome of the game by unilaterally changing his strategy.
In~other terms, in a~Nash equilibrium, each individual player has a
satisfactory strategy. Notice that Nash equilibria need not exist or be
unique, and are not necessarily optimal: Nash equilibria where all players
lose may coexist with more interesting Nash equilibria. Finding constrained
Nash equilibria (\eg, equilibria in which some players are required to win) is
thus an interesting problem for our setting.

\paragraph{Networks of identical devices.}
\looseness=-1
Our aim in this paper is to handle the special case where all the
interacting systems (but possibly a few of them) are
identical. This encompasses many situations involving computerized
systems over a network.
We propose a convenient way of
modelling such situations, and develop algorithms for synthesizing a
single strategy that, when followed by all the players, leads to a
global Nash equilibrium.
To~be meaningful, this requires symmetry assumptions on the
arena of the game (the board should look the same to all the players).
We also include \emph{imperfect observation} of the other players, 
which we believe is relevant in such a setting.

\looseness=-1
\paragraph{Our contributions.}  
We propose a convenient model for representing large interacting
systems, which we call \emph{game structure}. A~game structure is made of multiple
copies of a single arena (one copy per player); each player plays
on his own copy of the arena.
As mentioned earlier,
the players have imperfect information about
the global state of the game (they may have a perfect view on some of
their ``neighbours'', but may be blind to some other players).  In
\emph{symmetric} game structures, we~additionally require that any two
players are in similar situations: for every pair of players
$(A,B)$, we are able to map each player~$C$ to a corresponding
player~$D$ with the informal meaning that `player~$D$ is to~$B$
what player~$C$ is to~$A$'. Of~course, winning conditions and
imperfect information should respect that symmetry. We present several
examples illustrating the model, and argue why it is a relevant model
for computing symmetric Nash equilibria.

\looseness=-1
We show several undecidability results,  in
particular that the parameterized synthesis problem (aiming to
obtain one policy that forms a Nash equilibrium when applied to any
number of participants) is undecidable. We then
 characterize the complexity of computing (constrained) pure
symmetric Nash equilibria in symmetric game structures, when
objectives are given as \LTL formulas, and when restricting to
memoryless and bounded-memory strategies. 
This problem with no memory bound is then proven undecidable.

\paragraph{Related work.}
Game theory has been a very active area since the 1940's, but its applications
to computer science \textit{via} graph games is quite recent. In that domain,
until recently more focus had been put on zero-sum
games~\cite{tark2005-Hen}. Some recent works have
considered multi-player non-zero-sum games, including the computation of
(constrained) equilibria in turn-based and in concurrent
games~\cite{csl2004-CMJ,UW-lmcs11,BBMU-fossacs12}
or the development of temporal logics geared towards 
non-zero-sum objectives~\cite{concur2007-CHP,fsttcs2010-DLM}. 

None of those works distinguish symmetry constraints in strategy
profiles nor in game description.
Still, symmetry has been studied in the context
of normal-form games~\cite{Nas51,DM86}: in~such a
game, each player has the same set of actions, and the utility
function of a player only depends on his own action and on the number
of players who played each action (it~is independent on `who played
what').
Finally, let us mention that symmetry was also studied in the context of model
checking, where different techniques have been developped to deal with several
copies of the same system~\cite{GS92,ES96,AJ99}.

By lack of space, most of the technical developments could not be
included in this extended abstract. They are available in the
technical report~\cite{RR-lsv}.

\section{Symmetric concurrent games}

\subsection{Definitions}

\looseness=-1
For any~$k\in \bbN\cup\{\infty\}$, we~write $[k]$ for the set $\{i\in \bbN \mid 0\leq
i<k\}$ (in~particular, $[\infty]=\bbN$).
Let~$s=(p_i)_{i\in[n]}$ be a sequence, with $n\in\bbN\cup \{\infty\}$
being the length~$|s|$ of~$s$.
Let~$j\in\bbN$ s.t. $j-1<n$. The~$j$th element of~$s$, denoted~$s_{j-1}$, is the
element~$p_{j-1}$ (so that a~sequence~$(p_i)_{i\in[n]}$ may be named~$p$
when no ambiguity arises). The~$j$th prefix $s_{<j}$ of~$s$ is the finite
sequence $(p_i)_{i\in[j]}$. If~$s$ is finite, we~write $\last(s)$ for
its last element~$s_{|s|-1}$.  

\begin{definition}
An~\NMnewdef{arena} is a tuple $\tuple{\States, \Agt, \Act, \Mov, \Tab}$ where
\States is a finite set of states;
\Agt is a finite set of agents (also named players);
\Act is a finite set of actions;
$\Mov \colon \States\times \Agt \rightarrow
  2^{\Act} \setminus \{\emptyset\}$ is the set of actions available to
  a given player in a given state;
$\Tab \colon \States\times\Act^{\Agt}\rightarrow \States$ is a transition
  function that specifies the next state, given a state and an action of each
  player.
\end{definition}

The evolution of such a game is as usual: at each step, the players propose a
move, and the successor state is given by looking up this action vector in the
transition table. A~path is a sequence of states obtained this way; we~write
$\Hist$ for the set of finite paths (or~\emph{histories}).

Let~$A\in\Agt$. A~\NMnewdef{strategy} for~$A$ is a mapping~$\sigma_A\colon \Hist
\to \Act$ such that for any~$\rho\in\Hist$, $\sigma_A(\rho)
\in\Mov(\last(\rho),A)$. Given a set of players~$C\subseteq \Agt$, a~strategy
for~$C$ is a mapping $\sigma$ assigning to each~$A\in C$ a strategy
for~$A$ (we~write~$\sigma_A$ instead of $\sigma(A)$ to alleviate
notations).
As~a special case, a~strategy for~$\Agt$ is called a \NMnewdef{strategy
  profile}.
A~path~$\pi$ is \NMnewdef{compatible} with a strategy~$\sigma$ of coalition~$C$
if, for any~$i<|\pi|$, there exists a move~$(m_A)_{A\in\Agt}$ such that
$\Tab(\rho_{i-1}, (m_A)_{A\in\Agt})=\rho_{i}$ and $m_A=\sigma_A(\rho_{<i})$
for all~$A\in C$. The set of \NMnewdef{outcomes} of~$\sigma$ from a state~$s$,
denoted~$\Out(s,\sigma)$, is the set of plays from~$s$ that are compatible
with~$\sigma$.

Let~$\calG$ be a game. A~\NMnewdef{winning condition} for player~$A$ is
a set~$\Omega_A$ of plays of~$\calG$. We say that a play $\rho \in
\Omega_A$ yields payoff~$1$ to~$A$, and a play $\rho \notin \Omega_A$
yields payoff~$0$ to~$A$. 
A~strategy~$\sigma$ of a coalition~$C$ is \NMnewdef{winning} for~$A$ from a
state~$s$ if ${\Out(s,\sigma) \subseteq \Omega_A}$. 
A~strategy profile~$\sigma$ is a \NMnewdef{Nash equilibrium} if, for any~$A\in\Agt$ and
any strategy~$\sigma'_A$, if~$\sigma$ is losing for~$A$, then so
is~$\sigma[A\mapsto \sigma'_A]$. In~other terms, no~player can individually
improve his payoff. 

\begin{remark}
  In this paper, we only use purely boolean winning conditions, but
  our algorithms could easily be extended to the
  \emph{semi-quantitative} setting of~\cite{BBMU-fossacs12}, where
  each player has several (pre)ordered boolean objectives. 
  We omit such extensions in this paper, and keep focus on symmetry issues.
\end{remark}

\looseness=-1
The model we propose 
is made of a one-player arena, together with an observation
relation. Intuitively, each player plays in his own copy of the
one-player arena; the~global system is the product of all the local
copies, but each player observes the state of the global system only
through the observation relation. This is in particular needed for
representing large networks of systems, in which each player may only
observe some of his neighbours.

\begin{example}\label{ex-phone}
\looseness=-1
  Consider for instance a set of identical devices (\eg cell phones)
  connected on a local area network. Each device can modulate its
  emitting power. In~order to increase its bandwidth, a~device tends
  to increase its emitting power; but besides consuming more energy,
  this also adds \emph{noise} over the network, which decreases the other players' 
  bandwidth and encourages them to in turn increase their power.
  We~can model a device as an $m$-state
  arena (state~$i$ corresponding to some power~$p_i$, with $p_0=0$
  representing the device being~off). Any device would not
  know the exact state of the other devices, but would be able to
  evaluate the surrounding noise; this can be modelled using our
  observation relation. Based on this information, the device can
  decide whether it should increase or decrease its emitting power
  (knowing that the other devices play the same strategy as it is
  playing), resulting in a good balance between bandwidth and energy
  consumption. 
\end{example}

\begin{definition}
  An \emph{$n$-player game network} is a tuple $\mathcal{G} =
  \tuple{G,(\mathord\equiv_i)_{i \in [n]},(\Omega_i)_{i \in [n]}}$ s.t.
$G = \tuple{\States,\{A\},\Act,\Mov,\Tab}$ is a one-player arena;
for each $i \in [n]$, $\equiv_i$ is an equivalence relation
    on $\States^n$ (extended in a natural way to 
    sequences of states of~$\States^n$).
    Two $\equiv_i$-equivalent configurations are indistinguishable
    to player~$i$. This models \emph{imperfect information} for
    player~$i$;
for each $i \in [n]$, $\Omega_i \subseteq (\States^n)^\omega$
    is the objective of player~$i$. We~require that for all $\rho,\rho' \in
    (\States^n)^\omega$, if $\rho \equiv_i \rho'$ then $\rho$ and
    $\rho'$ are equivalently in~$\Omega_i$.
\end{definition}
The semantics of this game is defined as the ``product game'' 
$\mathcal{G'} = \tuple{\States',[n],\penalty0 \Act,\Mov',\Tab',(\Omega_i)_{i \in
    [n]}}$ where
$\States' = \States^n$,
$\Mov'((s_0,\dots,s_{n-1}),i) = \Mov(s,i)$,
and the transition table is defined as
\[
\Tab'((s_0,\dots,s_{n-1}),(m_i)_{i \in [n]}) = 
  (\Tab(s_0,m_0),\dots,\Tab(s_{n-1},m_{n-1})).
\]

An element of $\States^n$ is called a \NMnewdef{configuration} of
$\mathcal{G}$.  The equivalence relation~$\equiv_i$ induces equivalence
classes of configurations that player~$i$ cannot distinguish. We call
these equivalence classes \NMnewdef{information sets} and denote
$\mathcal{I}_i$ the set of information sets for player~$i$. Strategies
should respect these information sets: a~strategy~$\sigma_i$ for
player~$i$ is \NMnewdef{$\equiv_i$-realisable} whenever for all $\rho, \rho'
\in \Hist$, $\rho \equiv_i
\rho'$ implies $\sigma_i(\rho)=\sigma_i(\rho')$.  A~strategy profile
$\sigma=(\sigma_i)_{1 \le i \le n}$ is said \NMnewdef{realisable in~$\calG$}
whenever $\sigma_i$ is $\equiv_i$-realisable for every ${i \in [n]}$.

\begin{remark}
We assume that each equivalence relation $\equiv_i$ is given
\emph{compactly} using templates whose size is independant
of~$n$. As~an example, for~$P\subseteq \Agt$, the relation
$\textsf{Id}(P)$ defined by $(t,t')\in \textsf{Id}(P)$ iff
$t[i]=t'[i]$ for all~$i\in P$ encodes perfect observation of the
players in~$P$, and no information about the other players.
\end{remark}

\begin{example}
  Consider the cell-phone game again. It can be modelled as a game
  network where each player observes everything (\ie, the
  equivalence relations~$\equiv_i$ are the identity). A~more realistic
  model for the system can be obtained by assuming that each player
  only gets precise information about his close neighbours, and less
  precise information (only an estimation of the global noise in the network), or no
  information at all, about the devices that are far away.
\end{example}

\looseness=-1
Despite the global arena being described as a product of identical arenas, not
all games described this way are symmetric: the observation relation also
has to be \emph{symmetric}. We~impose extra conditions on that relation in order
to capture our expected notion of symmetry. 
Given a permutation~$\pi$ of~$[n]$, for a configuration $t = (s_i)_{i\in[n]}$
we~let $t(\pi) =
(s_{\pi(i)})_{i\in[n]}$; for a path $\rho = (t_j)_{j \in \bbN}$, 
we~let $\rho(\pi) = (t_j(\pi))_{j \in \bbN}$.

\begin{definition}
  \label{def:scgs}
  A game network $\mathcal{G} = \tuple{G,(\mathord\equiv_i)_{i \in
    [n]},\penalty100 (\Omega_i)_{i \in [n]}}$ is \emph{symmetric}
  whenever for any two players~$i,j \in [n]$, there is a
  permutation~$\pi_{i,j}$ of~$[n]$ such that $\pi_{i,j}(i)=j$ and 
  satisfying the following conditions for every $i,j,k \in [n]$:
  \begin{enumerate}
  \item\label{sym1} $\pi_{i,i}$ is the identity, and $\pi_{k,j} \circ \pi_{i,k}=
    \pi_{i,j}$; hence $\pi_{i,j}^{-1}=\pi_{j,i}$.
  \item\label{sym2} the observation made by the players is compatible with the
    symmetry of the game: 
    for any two configurations $t$ and~$t'$, $t \equiv_i t'$ iff
      $t(\pi_{i,j}^{-1}) \equiv_j t'(\pi_{i,j}^{-1})$;
    \item\label{sym3} objectives are compatible with the symmetry of
      the game: for every play~$\rho$, $\rho \in \Omega_i$ iff
      $\rho(\pi_{i,j}^{-1}) \in \Omega_j$.
  \end{enumerate}
  In that case, $\pi = (\pi_{i,j})_{i,j \in [n]}$ is called a
  \emph{symmetric representation} of~$\mathcal{G}$.
\end{definition}

\looseness=-1 
The mappings $\pi_{i,j}$ define the symmetry of the game:
$\pi_{i,j}(k)=l$ means that player~$l$ plays vis-\`a-vis player~$j$
the role that player~$k$ plays vis-\`a-vis player~$i$.  We~give the
intuition why we apply~$\pi_{i,j}^{-1}$ in the definition above, and
not~$\pi_{i,j}$.
Assume configuration $t=(s_0,\dots,s_{n-1})$ is observed by player
$i$. The corresponding configuration for player $j$ is
$t'=(s'_0,\dots,s'_{n-1})$ where player-$\pi_{i,j}(k)$ state should be
that of player~$k$ in~$t$. That is, $s'_{\pi_{i,j}(k)}=s_k$, so~that
$t' = t(\pi_{i,j}^{-1})$.

These mappings also define how symmetry must be used in strategies: let
$\mathcal{G}$ be a symmetric $n$-player game network with symmetric
representation~$\pi$. We~say that a strategy profile $\sigma=(\sigma_i)_{i \in
  [n]}$ is \emph{symmetric} for the representation~$\pi$ if it is realisable
(\ie, each player only plays according to what he can observe) and if for all
$i,j \in [n]$ and every history~$\rho$, it~holds
\(
\sigma_i(\rho) = \sigma_j(\rho(\pi_{i,j}^{-1}))
\).

The following lemma characterizes symmetric strategy profiles:
\begin{restatable}{lemma}{stratprofile}
  \label{lemma:symmetry}
  Fix a symmetric representation $\pi$ for $\mathcal{G}$. If
  $\sigma_0$ is an $\equiv_0$-realisable strategy for player~$0$, then
  the strategy profile~$\sigma$ defined for all~$i>0$ by $\sigma_i(\rho) =
  \sigma_0(\rho(\pi_{i,0}^{-1}))$ is symmetric.
\end{restatable}

\begin{example}
  Consider a card game tournament with six players, three on each
  table. Here each player has a left neighbour, a right neighbour, and
  three opponents at a different table.
  To model this, one could assume player~$0$ knows everything about
  himself, and has some informations about his right neighbour (player~$1$)
  and his left neighbour (player~$2$). But he knows nothing about players~$3$,
  $4$ and~$5$.

  Now, the role of player~$2$ vis-\`a-vis player~$1$ is that of
  player~$1$ vis-\`a-vis player~$0$ (he is his right
  neighbour). Hence, we can define the symmetry as $\pi_{0,1}(0)=1$,
  $\pi_{0,1}(1)=2$, $\pi_{0,1}(2)=0$, and
  $\pi_{0,1}(\{3,4,5\})=\{3,4,5\}$ (any choice is fine here).  As~an
  example, the observation relation in this setting could be that
  player~$0$ has perfect knowledge of his set of cards, but only knows
  the number of cards of players~$1$ and~$2$, and has no information
  about the other three players. Notice that other observation
  relations would have been possible (for instance, giving more
  information about the right player).
\end{example}

In this paper we are interested in computing (symmetric) Nash
equilibria in symmetric game networks:

\begin{problem}[Constrained existence of (symmetric) NE]
\looseness=-1
  The \emph{constrained existence problem} asks, given a symmetric
  game network $\mathcal{G}$, a~symmetric representation~$\pi$, a
  configuration~$t$, a~set $L \subseteq [n]$ of losing players, and a
  set $W \subseteq [n]$ of winning players, whether there is a
  (symmetric) Nash equilibrium~$\sigma$ in~$\mathcal{G}$ from~$t$ for the
  representation~$\pi$,
  such that all players in~$L$ lose and all players in~$W$~win.  
  If~$L$ and~$W$ are empty, the problem is simply called the \emph{existence
    problem}. 
  If $W = [n]$, the problem is called the \emph{positive existence problem}.
\end{problem}

We first realise that even though symmetric Nash equilibria are Nash
equilibria with special properties, they are in some sense at least as
hard to find as Nash equilibria.
This can be proved by seeing the individual game structure as a
product of $n$ disconnected copies of the original individual
structure. This~way, the~strategy played by one player on one copy
imposes no constraints on the strategy played by another player on a
different copy.
\begin{restatable}{proposition}{existtosym}
  \label{prop:existtosym}
  From a symmetric game network $\mathcal{G}$ we can construct in
  polynomial time a symmetric game network~$\mathcal{H}$ such that
  there exists a symmetric Nash equilibrium in~$\mathcal{H}$ if, and
  only if, there exists a Nash equilibrium
  in~$\mathcal{G}$. Furthermore the construction only changes the
  arena, but does not change the number of players nor the objectives
  or the resulting payoffs.
\end{restatable}


\section{Our results}
\paragraph{Undecidability with non-regular objectives.}
\looseness=-1 
Our games allow for arbitrary boolean objectives, defined for each
player as a set of winning plays. As can be expected, this is too
general to get decidability of our problems even with perfect
information, since it can be used to encode the runs of a two-counter machine:
\begin{restatable}{theorem}{theoundec}
  \label{theo:undec2}
  The (constrained)
  existence of a symmetric Nash equilibrium for
  non-regular objectives in (two-player) perfect-information symmetric game networks
  is undecidable.
\end{restatable}

\paragraph{Undecidability with a parameterized number of players.}

\looseness=-1 
Parameterized synthesis of Nash equilibria (that is,
synthesizing a single strategy that each player will apply, and that yields a
Nash equilibrium for any number of players) was one of our targets
in this work. We show that computing such equilibria is not possible,
even in rather restricted settings.

\begin{restatable}{theorem}{undecparam}
  \label{theo:undecparam}
  The (positive) existence of a \emph{parameterized} symmetric Nash
  equilibrium for \LTL objectives in symmetric game
  networks is undecidable (even for memoryless strategies).
\end{restatable}
This is proved by encoding a Turing machine as a game network with
arbitrarily many players, each player controlling one cell of the
tape. The machine halts if there exists a number~$n$ of players such
that the play reaches the halting state. We use \LTL formulas to
enforce correct simulation of the Turing machine.

\paragraph{From positive existence to existence.}
\looseness=-1
Because of the previous result, we now fix the number~$n$ of players. 
Before turning to our decidability results, 
we begin with showing that 
positive existence of Nash equilibria is not harder than existence.
Notice that this makes a difference with
the setting of turn-based games, where Nash equilibria always exist.

\begin{restatable}{proposition}{cextoex}
  \label{prop:cextoex}
  Deciding the (symmetric) existence problem in (symmetric) game
  networks is always at least as hard as deciding the positive
  (symmetric) existence problem. The reduction doubles the number of
  players and uses \LTL objectives, but does not change the nature of
  the strategies (memoryless, bounded-memory, or general).
\end{restatable}

\paragraph{Bounded-memory strategies.}

\begin{theorem}
  \label{thm:finite-memory}
  The (positive, constrained) existence of a bounded-memory symmetric Nash
  equilibrium for \LTL objectives in symmetric game
  networks is \EXPSPACE-complete.
\end{theorem}

The \EXPSPACE-hardness results are direct consequences of the proof of
Theorem~\ref{theo:undecparam} (the only difference is that we restrict
to a Turing machine using exponential space).

The algorithm for memoryless strategies is as follows: it~first
guesses a memoryless strategy for one player, from which it deduces
the strategy to be played by the other players. It~then looks for the
players that are losing, and checks if they alone can improve their
payoff. If they cannot improve the guessed strategy yields a Nash
equilibrium, otherwise it does not yield an equilibrium.

The first step is to guess and store an $\equiv_0$-realisable
memoryless strategy~$\sigma_0$ for player~$0$, which we then
prove witnesses the existence of a symmetric Nash equilibrium. Such a
strategy is a mapping from $\States^n$ to~$\Act$. We~intend player~$0$
to play according to~$\sigma_0$, and any player~$i$ to play according
to $\sigma_0(\pi_{i,0}^{-1}(s_0,...,s_{n-1}))$ in
state~$(s_0,...,s_{n-1})$. From Lemma~\ref{lemma:symmetry} we know
that all symmetric memoryless strategy profiles can be characterized
by such an $\equiv_0$-realisable memoryless strategy for
player~$0$.

The algorithm then guesses a set~$W$ of players (which satisfies the
given constraint), and checks that under the strategy profile computed
above, the players in~$W$ achieve their objectives while the players
not in~$W$ do~not. This is achieved by computing the non-deterministic
B\"uchi automata for~$\phi_i$ if~$i\in W$ and for $\neg\phi_i$
if~$i\notin W$, and checking that the outcome of the strategy profile
above (which is a lasso-shaped path and can easily be computed from
strategy~$\sigma_0$) is accepted by all those automata.

\looseness=-1
It~remains to check that the players not in~$W$ cannot win if they
deviate from their assigned strategy. For each player~$i$ not in~$W$,
we~build the one-player game where all players but player~$i$ play
according to the selected strategy profile. The resulting automaton
contains all the plays that can be obtained by a deviation of
player~$i$. It~just remains to check that there is no path satisfying
$\phi_i$ in that automaton. If this is true for all players not
in~$W$, then the selected strategy~$\sigma_0$ gives rise to a
memoryless symmetric Nash equilibrium.

Regarding (space) complexity, storing the guessed strategy requires
space $O(|\States|^n)$. The B\"uchi automata have size exponential in
the size of the formulas, but can be handled on-the-fly using
classical constructions, so that the algorithm only requires
polynomial space in the size of the formula. The lasso-shaped outcome,
as well as the automata representing the deviations of the losing
players, have size $O(|\States|^n)$, but can also be handled
on-the-fly. In~the end, the whole algorithm runs in exponential space
in the number of players, and polynomial in the size of the game and 
in the size of the \LTL formulas.

\smallskip

The above algorithm can be lifted to bounded-memory
strategies: given a memory bound~$m$, it~guesses a strategy~$\sigma_0$
using memory~$m$, and does the same computations as above. Storing the
strategy now requires space $O(m\cdot |\States|^n)$, which is still exponential, 
even if $m$ is given in binary.

\begin{remark}
  Notice that the algorithms above could be adapted to handle
  non-symmetric equilibria in non-symmetric game networks: it~would
  just guess all the strategies, the payoff, and check the
  satisfaction of the \LTL objectives in the product automaton
  obtained by applying the strategies.

\looseness=-1
  The algorithm could also be adapted, still with the same complexity, to handle richer
  objectives, in particular in the semi-quantitative setting
  of~\cite{BBMU-fossacs12}, where the players have several (pre)ordered
  objectives. Instead of guessing the set of winners, the algorithm  would guess, for each
  player, which objectives are satisfied, and check that no individual
  improvement is possible. The latter can be achieved by listing all possible
  improvements and checking that none of them can be reached.
\end{remark}

\paragraph{General strategies.}

\looseness=-1 We already mentioned an undecidability result in
Theorem~\ref{theo:undec2} for two-player games and perfect information when
general strategies are allowed.  However, the objectives used for
achieving the reduction are quite complex. On the other hand,
imperfect information also leads to undecidability for \LTL objectives
with only 3 players. To show this, we can slightly alter a proof
from~\cite{PR90}. Here, synthesis of distributed
reactive systems (corresponding to finding sure-winning strategies)
with \LTL objectives is shown undecidable in the presence of imperfect
information. The situation used in the proof can be modelled in our
framework and by adding a matching-penny module in the beginning and
slightly changing the \LTL objectives, we can obtain undecidability of
Nash equilibria instead of sure-winning strategies.

\begin{restatable}{theorem}{undecimperfect}
  \label{theo:undecimp}
  The existence of a (symmetric)
  Nash equilibrium for \LTL objectives in symmetric game
  networks is undecidable for $n \ge 3$ players.
\end{restatable}


\enlargethispage{3mm}

\end{document}